\documentclass[10pt]{article}
\usepackage[BOE]{express}
\usepackage{subfig}
\usepackage[font=footnotesize]{caption}
\usepackage{amsmath,amssymb}
\def \rs{\mathbf{r}_s}
\def \r{\mathbf{r}}
\def \rd{\mathbf{r}_d}
\def \gex{g_{ex}}
\def \gem{g_{em}}
\def \PHI{\mathbf{\Phi}}
\def \G{\mathbf{G}}
\def \x{\mathbf{x}}
\def \n{\mathbf{n}}
\def \U{\mathbf{U}}
\def \SIGMA{\mathbf{\Sigma}}
\def \V{\mathbf{V}}
\def \M{\mathbf{M}}

\def \EPSILON{\mathbf{\epsilon}}

\def \y{\mathbf{y}}
\def \A{\mathbf{A}}
\begin{document}

\title{Image reconstruction in fluorescence molecular tomography with sparsity-initialized maximum-likelihood expectation maximization}

\author{Yansong Zhu,\authormark{1,2} Abhinav K. Jha,\authormark{2,3,4,*} Dean F. Wong,\authormark{2,5,6,7} and Arman Rahmim\authormark{1,2}}

\address{\authormark{1}Department of Electrical and Computer Engineering, Johns Hopkins University, Baltimore, MD, USA\\
\authormark{2}Department of Radiology and Radiological Science, Johns Hopkins University, Baltimore, MD, USA\\
\authormark{3}Department of Biomedical Engineering, Washington University in St. Louis, St. Louis, MO, USA\\
\authormark{4}Mallinckrodt Institute of Radiology, Washington University in St. Louis, St. Louis, MO, USA\\
\authormark{5}Department of Neuroscience, Johns Hopkins University, Baltimore, MD, USA\\
\authormark{6}Department of Psychiatry and Behavioral Science, Johns Hopkins University, Baltimore, MD, USA\\
\authormark{7}Department of Neurology, Johns Hopkins University, Baltimore, MD, USA\\
}

\email{\authormark{*}ajha4@jhmi.edu} %% email address is required



\begin{abstract}
We present a reconstruction method involving maximum-likelihood expectation maximization (MLEM) to model Poisson noise as applied to fluorescence molecular tomography (FMT). MLEM is initialized with the output from a sparse reconstruction-based approach, which performs truncated singular value decomposition-based preconditioning followed by fast iterative shrinkage-thresholding algorithm (FISTA) to enforce sparsity. The motivation for this approach is that sparsity information could be accounted for within the initialization, while MLEM would accurately model Poisson noise in the FMT system. Simulation experiments show the proposed method significantly improves images qualitatively and quantitatively. The method results in over 20 times faster convergence compared to uniformly initialized MLEM and improves robustness to noise compared to pure sparse reconstruction.
We also theoretically justify the ability of the proposed approach to reduce noise in the background region compared to pure sparse reconstruction. Overall, these results provide strong evidence to model Poisson noise in FMT reconstruction and for application of the proposed reconstruction framework to FMT imaging.
\end{abstract}

 % REPLACE WITH CORRECT OCIS CODES FOR YOUR ARTICLE, MINIMUM OF TWO; Avoid using the OCIS codes for “General” or “General science” whenever possible.
%For a complete list of OCIS codes, visit: https://www.osapublishing.org/oe/submit/ocis/

%%%%%%%%%%%%%%%%%%%%%%% References %%%%%%%%%%%%%%%%%%%%%%%%%

%%%%%%%%%%%%%%%%%%%%%%%%%%  body  %%%%%%%%%%%%%%%%%%%%%%%%%%
\section{Introduction}
%Fluorescence molecular tomography (FMT) is a method that provides 3D visualization and quantification of the distribution of molecular target within biological tissue. Photon propagation in biological tissue suffers from high degree of absorption and scattering, and number of measurement acquired at tissue boundary is usually far less compared to the number of unknown parameters. These make the inverse problem of FMT an ill-posed and underdetermined problem. To get a stable solution, appropriate prior knowledge is assuemed before the reconstruction. The fluorescence molecules tends to concentrate at target region, which makes the sparse assumption a proper prior knowledge. In recent years, compressed sensing (CS) technique has been explored for reconstruction problem in FMT. Condition of incoherence required for CS technique is satisfied with certain pre-conditioning technique. 
Fluorescence molecular tomography (FMT) is finding several applications in 3D visualization and quantification of the distribution of molecular target within biological tissue\cite{ntziachristos2006fluorescence}.
In particular, FMT has received substantial interest in small animal imaging for applications such as studying tumor physiology and for pharmaceutical research\cite{biswal2011imaging,stuker2011fluorescence}.
In FMT imaging, fluorescence molecules are first injected into biological tissue. External illumination sources are used to excite the fluorescence molecules. The photons emitted by the excited fluorescence molecules are collected by detectors at the tissue surface. The objective in FMT is to use these surface measurements to reconstruct the 3D distribution of fluorescence molecules within the tissue.

The reconstruction problem in FMT is known to be highly ill-posed, and is sensitive to noise and modeling errors such as discretization\cite{Zhou2011discre,Zhao:14}. Over the past two decades, various reconstruction methods for FMT have been proposed\cite{arridge2009optical}. Tikhonov regularization is a popular regularization applied to FMT reconstruction problem. The regularized problem can be solved iteratively with methods such as Newton method and algebraic reconstruction technique (ART)\cite{arridge2009optical,ntziachristos2001}. However, such regularization tends to over-smooth the reconstructed images, leading to loss of localized features during reconstruction\cite{ye2014fast}.
More recently, reconstruction methods that exploit sparsity of the fluorescence distribution have been studied\cite{Han:10,Zhao:14,dutta2012joint,lee2011compressive}. In these methods, $\ell_0$ or $\ell_1$ regularization on the fluorescence distribution is applied to enforce sparsity while performing the reconstruction. These regularization problems can be solved with methods such as greedy algorithms and iterative thresholding methods\cite{bruckstein2009sparse}.

The noise in the data acquired using FMT systems is Poisson distributed\cite{MP:MP0209}. For this noise distribution, MLEM-based reconstruction techniques have yielded reliable results, especially in nuclear medicine imaging
 \cite{vardi1985statistical,lange1987theoretical,llacer1993statistical,lewitt2003overview,jha2013joint}.
The MLEM technique has several advantages, such as accurately modeling the Poisson noise distribution in the acquired data, constraining the activity values to be non-negative without the need for a specific regularizer, and ensuring the conservation of the total number of photons across multiple iterations.
In optical tomography, several studies have applied MLEM for reconstruction in bioluminescence tomography\cite{Alexandrakis2005,slavine2006iterative,jiang2007image}. In \cite{cao2010bayesian}, MLEM has also been applied for FMT reconstruction.
However, the MLEM technique typically suffers from slow convergence for optical tomography modalities, with thousands of iterations and large amount of time per iteration being required\cite{qi2006iterative,ahn2008fast,jiang2007image}. This makes MLEM a time-consuming method and thus not very practical\cite{Alexandrakis2005,cao2010bayesian}. As a result, MLEM has not been widely used for image reconstruction in optical tomography.

The performance of MLEM is influenced by different factors. An important factor being the initial estimate provided to the algorithm. Conventionally, MLEM starts with a uniform initial estimate, as we explain later. However, different initializations for MLEM yield different reconstruction results\cite{Barrett:04,ma2013performance}.
In this work, we studied the use of sparse reconstruction to initialize the MLEM approach. The overall motivation for this approach is that the sparse reconstruction method would account for the sparsity of the fluorescence distribution, while the MLEM would accurately model the Poisson noise in the FMT system. However, this combined approach is also able to exploit several inherent advantages of these two techniques, as we describe below. Our method yield reliable and improved results in comparison to pure sparse reconstruction as well as uniformly initialized MLEM methods. Preliminary versions of this work have been presented previously\cite{zhu:17,Zhu:18}. We begin by describing our method in the next section.
\section{Methods}
\subsection{The forward model and reconstruction problem in FMT}
The forward model in FMT is described by a pair of coupled equations. The first equation describes the propagation of excitation photons from source at location $\rs$ to location $\r$ in the medium and the second one describes the propagation of emitted fluorescence photons from location $\r$ to detector at location $\rd$, where $\r_s$, $\r$ and $\r_d$ are three-dimensional vectors. These coupled equations are given by:
\begin{equation}
\phi_{ex}(\r)=\int_{\Omega}{\gex(\rs, \r)s(\rs)d\rs},
\end{equation}
and 
\begin{equation}
\phi_{em}(\rd)=\int_{\Omega}{\gem(\r, \rd)x(\r)\phi_{ex}(\r)d\r},
\end{equation}
where $\phi_{ex}(\r)$ and $\phi_{em}(\r_d)$ are the excitation light field at location $\r$ and emission light field at detector location $\r_d$, respectively, $\gex(\rs, \r)$ is the Green's function of excitation light at location $\r$ due to a source at location $\rs$, $\gem(\r, \rd)$ denotes Green's function of emission light detected by detector at location $\rd$ due to the fluorescence source at location $\r$, $x(\r)$ is the fluorescence yield at location $\r$, and $\Omega$ denotes object support,.
If we discretize $\Omega$ into $N$ voxels, we obtain the linear matrix equation for the forward model:
\begin{equation}
\PHI=\G \x,
\label{fwd}
\end{equation}
where
\begin{equation*}
\G=
\begin{bmatrix}
g_{em,1}^1\phi_{em,1}^1 & \dots & g_{em,N}^1\phi_{em,N}^1\\
\vdots 							&		 &\vdots 							\\
g_{em,1}^{N_d}\phi_{em,1}^1 & \dots & g_{em,N}^{N_d}\phi_{em,N}^1 \\
g_{em,1}^1\phi_{em,1}^2 & \dots & g_{em,N}^1\phi_{em,N}^2\\
\vdots 							&		 &\vdots 							\\
g_{em,1}^{N_d}\phi_{em,1}^{N_s} & \dots & g_{em,N}^{N_d}\phi_{em,N}^{N_s} \\
\end{bmatrix}
\end{equation*}
is the sensitivity matrix of the system, $\PHI$ is an $M\times 1$ vector denoting detector measurements, $\x$ is an $N\times 1$ vector representing unknown fluorescence yield, $N_s$ and $N_d$ are number of sources and detectors, respectively, and $M=N_s\times N_d$ is the total number of measurements. Due to the limited number of sources and detectors, typically $M<N$ in FMT.

%Based on the forward model, the goal of FMT is to reconstruct $\x$ given the measurements $\PHI$ and sensing matrix $\G$. 
Modeling the measurement noise denoted by the $M$-dimensional vector $\n$, Eq. (\ref{fwd}) becomes:
\begin{equation}
\PHI=\G \x+\n.
\label{fwdn}
\end{equation}
In FMT, the data collected by the detectors is corrupted by Poisson noise\cite{MP:MP0209}.
The reconstruction problem in FMT is to reconstruct $\x$ given sensitivity matrix $\G$ and detector measurements $\PHI$. 
In the next section, we derive the MLEM-based reconstruction technique that models this noise distribution accurately.
\subsection{Modeling Poisson noise in the reconstruction}
The likelihood function for Poisson distributed data is:
\begin{equation}
l(\x|\PHI)=\prod_{m=1}^M{\exp\left[-(\G\x)_m\right]}\frac{(\G\x)_m^{\phi_m}}{\phi_m!},
\end{equation}
where $(\G\x)_m$ and $\phi_m$ denote the $m^{th}$ element of the vector $(\G\x)$ and $\PHI$, respectively. Taking the logarithm of the likelihood function yields:
\begin{equation}
L(\x|\PHI)=\sum_{m=1}^M\left\{-(\G\x)_m+\phi_m\ln\left[(\G\x)_m\right]-\ln\phi_m!\right\}.
\end{equation}
The first order derivative of the log-likelihood function is given by
\begin{equation}
\frac{\partial}{\partial x_n}L(\x|\PHI)=\sum_{m=1}^M\left\{-G_{mn}+\frac{\phi_m}{(\G\x)_m}G_{mn}\right\}.
\end{equation}
Setting $\frac{\partial}{\partial x_n}L(\x|\PHI)=0$ yields
\begin{equation}
1=\frac{1}{\sum_{m=1}^M{G_{mn}}}\sum_{m=1}^M{\frac{\phi_m}{(\G\x)_m}G_{mn}}.
\end{equation}
Multiplying both side with $x$ and replacing $x$ with a sequence of estimates $\hat{x}^k$ yields the fixed-point iteration:
\begin{equation}
\hat{x}_n^{k+1}=\hat{x}_n^{k}\frac{1}{s_n}\sum_{m=1}^M{\frac{\phi_m}{(\G \hat{\x}^k)_m}G_{mn}},
\label{equ:mlem}
\end{equation}
where $s_n=\sum_{m=1}^M{G_{mn}}$. This is referred to the MLEM technique\cite{Barrett:04}.

The MLEM iteration starts from an initial estimate $\hat{\x}^0$, and the results of this technique can be influenced by its initial estimate\cite{Barrett:04}. Typically, the initial estimate is uniform, where all the elements in $\hat{\x}^0$ are assumed to be a constant\cite{kontaxakis1998maximum,chang2004regularized}. However, with this estimate, MLEM updates all the voxels in every iteration, increasing the computational requirements. In Eq. (\ref{equ:mlem}), note that $\hat{x}^k_n$ will always be zero if $\hat{x}^0_n=0$ due to the multiplicative nature of the technique. Thus, the zero elements can be excluded from $\hat{\x}^0$ during MLEM iteration. Matrix $\G$ used for MLEM iteration can be formulated with columns corresponding to non-zero elements in $\hat{\x}^0$. This reduces the size of matrices in the reconstruction problem and accelerates the computation speed. In this context, in many FMT applications, fluorescence molecules tend to concentrate in a small target region. Thus, if we could exploit this property, we could generate a sparse initial estimate, which allows us to accelerate the MLEM technique. Such a technique would inherently exploit the sparsity-based prior information in FMT as well as model the Poisson noise in FMT accurately. Inspired by this, we developed a sparse reconstruction method and used the output from this method as the initial estimate for MLEM. In the next section, we describe the method we used to obtain sparse initial estimate of MLEM. 
%Since zero elements won't be updated in MLEM, system matrix $\G$ is formulated with columns corresponding to non-zero elements in $\x$. The negative elements in result of sparse reconstruction are set to zero because of the non-negative constraints of MLEM.  

\subsection{Sparse reconstruction and preconditioning of sensitivity matrix}
To provide the sparse initial estimate for MLEM, the following minimization problem can be formulated based on Eq. (\ref{fwdn}):
\begin{equation}
\begin{aligned}
\min_{\x}{\|\x\|_0}     & &      \text{such that}     & &    \|\mathbf{\PHI}-\mathbf{\G\x}\|_2\leq\epsilon.
\end{aligned}
\label{equ:l0}
\end{equation}
While directly solving this problem is computationally complex, Eq. (\ref{equ:l0}) can be approximately solved with greedy algorithms or convex relaxation techniques\cite{bruckstein2009sparse}. The theory of compressed sensing (CS) provides the conditions under which such approximate solvers are valid. Further, approaches based on singular value decomposition (SVD) can be applied to the sensitivity matrix to improve sparse reconstruction in FMT\cite{jin2012preconditioning,shi2013greedy,jin2014light,yao2015wide}. This technique is known as preconditioning of sensitivity matrix.
Here, we follow truncated singular value decomposition (TSVD) described in \cite{shi2013greedy} as the preconditioning method. First, expressing the matrix $\G$ in terms of its singular vectors and singular values using SVD, Eq. (\ref{fwdn}) becomes:
\begin{equation}
\PHI=\U \SIGMA \V^T \x+\n,
\label{svd}
\end{equation}
where $\U$ and $\V$ are $M\times M$ and $N\times N$ unitary matrices where the columns are left-singular vectors and right-singular vectors, respectively, and $\SIGMA$ is a diagonal matrix where the diagonal elements are the singular values. By multiplying both sides of Eq. (\ref{svd}) with $\SIGMA^{-1}\U^T$, we could potentially use $\V^T$ as the new sensitivity matrix. However, since the reconstruction problem in FMT is highly ill-posed, the inversion of small singular values contained in $\SIGMA$ will cause large noise amplification. To address this issue, we keep only the $K$ largest singular values of matrix $\SIGMA$ and discard the rest, before performing the inversion of $\SIGMA$. The corresponding columns in $\U$ and $\V$ are also discarded. This process is referred to as truncation. Then  Eq. (\ref{svd}) becomes 
\begin{equation}
\PHI=\U_t \SIGMA_t \V_t^T \x+\n,
\label{tsvd}
\end{equation}
where the size of $\U_t$, $\SIGMA_t$ and $\V_t$ are $M\times K$, $K\times K$ and $N\times K$, respectively. Since small singular values are discarded, usually $K<M$.
Applying $\M=\SIGMA_t^{-1}\U_t^T$ to both sides of Eq. (\ref{tsvd}) yields
\begin{equation}
\M \PHI=\V_t^T \x+\M \n.
\label{equ:mask}
\end{equation}
Denoting $\y=\M\PHI$, $\A=\V_t^T$ and $\n'=\M \n$, Eq. (\ref{equ:mask}) can be written as
\begin{equation}
\y=\A \x+\n'.
\label{equ:afterprecond}
\end{equation}
We now solve Eq.(\ref{equ:afterprecond}) as a sparse reconstruction problem. More specifically, we implemented convex relaxation technique in this work. Our objective is to minimize the $\ell_1$ norm of the vector $\x$. Thus the sparse reconstruction problem is posed as
\begin{equation}
\min_{\x}{\|\x\|_1} \quad   \textrm{such that} \quad  \|\y-\A\x\|_2\leq\epsilon.
\label{equ:sparse}
\end{equation}
We applied the fast iterative shrinkage-thresholding algorithm (FISTA) for solving the minimization problem in Eq. (\ref{equ:sparse})\cite{BeckFISTA}. The output with this method is then input to the MLEM technique as the initial estimate.
Note that results from sparse reconstruction might contain negative elements. As we explained previously, MLEM constrains the activity values to be non-negative. To enable this, the negative elements in $\hat{\x}^0$ are set to zero.
\subsection{Experiments}
\begin{figure}
\centering
\subfloat[]{\includegraphics[width=0.5\textwidth]{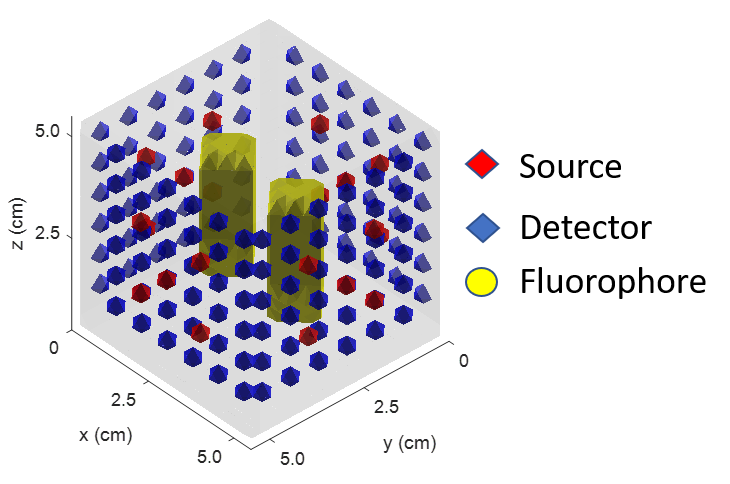}} \hspace{0.5 cm}
\subfloat[]{\includegraphics[width=0.4\textwidth]{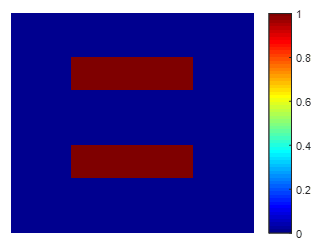}} \\
\caption{(a) The experimental setup of cube phantom. (b) Cross section at $y=2.5\textrm{ cm}$ of the simulated phantom. }
\label{fig:cube_setup}
\end{figure} 
To validate the proposed method, different simulation experiments were conducted. Three different reconstruction methods were implemented for comparison, namely, (a) a pure sparsity-based reconstruction method that used TSVD in conjunction with FISTA, (b) the MLEM method with uniform initial estimate of the image (more specifically, the initial activity values in all the voxels was set to unity) and (c) the MLEM method with an initialization that was obtained using the method described in (a). We will refer to these methods as pure sparsity-based reconstruction method, uniformly initialized MLEM and sparsity-initialized MLEM, respectively.
%MLEM with uniform initialization, and MLEM with sparse initialization. For sparse reconstruction, we implemented fast iterative shrinkage-thresholding algorithm (FISTA) to solve Eq. (\ref{equ:sparse}) after preconditioning\cite{BeckFISTA}.

In the first set of experiments, a $5\times 5\times 5\textrm{ cm}^3$ cubic phantom was considered, as shown in Fig.~\ref{fig:cube_setup}(a). The phantom was discretized into $20\times 20\times 20$ voxels. The absorption coefficient of the phantom was set to $\mu_a=0.05\textrm{ cm}^{-1}$ and the reduced scattering coefficient was set to $\mu_s'=10\textrm{ cm}^{-1}$. $20$ sources and $144$ detectors were placed on the side surfaces. This configuration generated $2880$ measurements. Two cylindric fluorescence bars with radius of $0.375\textrm{ cm}$ and length of $2.5\textrm{ cm}$ each were inserted into the phantom. The fluorescence intensity in these bars was set to unity. The cross section of the phantom at $y=2.5\textrm{ cm}$ is shown in Fig.~\ref{fig:cube_setup}(b). The Green's function in the forward model of FMT was computed using Monte Carlo method, where a large number of photons were simulated to generate approximately noiseless measurements\cite{Fang:09}. The measurements were then scaled to different levels and corresponding Poisson noise was applied using a Poisson distributed pseudo random number generator. This yielded detector measurements with different signal-to-noise ratio (SNR) values.
\begin{table}[ht!]
\centering
\caption{Optical properties of digital mouse phantom\cite{strangman2003factors}}
\begin{tabular}{c | c | c | c}
\hline
Tissue type & Brain & Skull & Skin \\
$\mu_s'(\textrm{ cm}^{-1})$ & 12.5 & 10.0 & 8.0  \\
$\mu_a(\textrm{ cm}^{-1})$  & 0.178 & 0.101 & 0.159\\
\hline
\end{tabular}
\label{table1}
\end{table}
\begin{figure}
\centering
\subfloat[]{\includegraphics[width=0.5\textwidth]{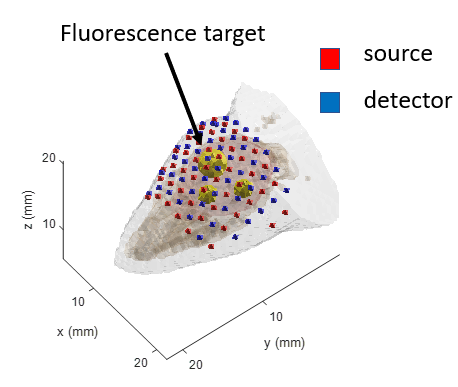}} \hspace{0.3 cm}
\subfloat[]{\includegraphics[width=0.4\textwidth]{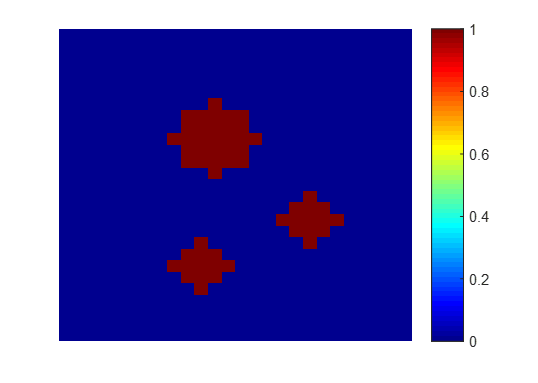}} \\
\caption{(a) The experimental setup of mouse phantom. (b) Cross section of digital mouse phantom at $z=16\textrm{mm}$.}
\label{fig:moby_phan}
\end{figure} 
\begin{figure}
\centering\includegraphics[width=\textwidth]{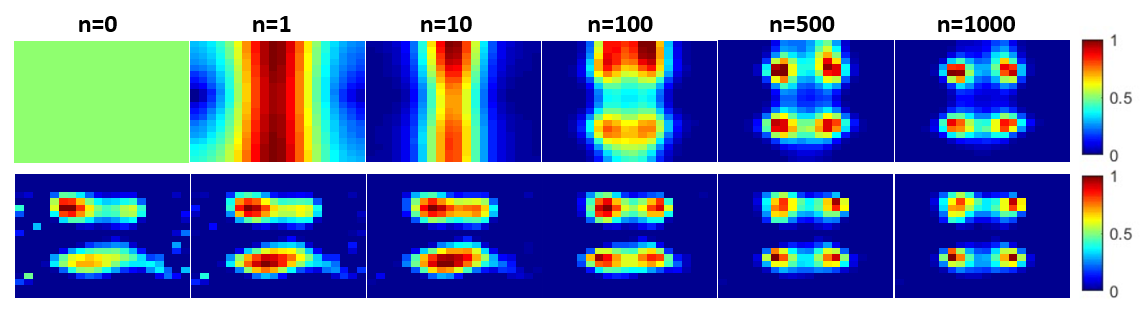}
\caption{Cross sections at $y=2.5\textrm{ cm}$ reconstructed by MLEM with different iteration number n for a cube phantom. SNR=$18$dB. The reconstructed images are from MLEM with uniform initial estimate for the top row and MLEM with sparse initial estimate for the bottom row.}
\label{fig:cube_crosec}
\end{figure}

To study the effect of MLEM iteration number on reconstruction performance, $1000$ iterations were performed for MLEM with different initializations with the SNR initially set to $18$dB, and the truncation number $K$ set to $760$. The region of interest (ROI) corresponded to the region occupied by the fluorescence bars. The rest of the region was defined as background. For quantitative study, different figures of merit were computed. Specifically, we computed absolute bias in the estimated uptake in the ROI and the background, spatial variance within the pixels in the ROI and the background, and the root mean square error (RMSE) for the entire image. The mean of the fluorescence uptake within the ROI, denoted by $\theta_{\textrm{ROI}}$, is defined as
\begin{figure}
\centering\includegraphics[width=\textwidth]{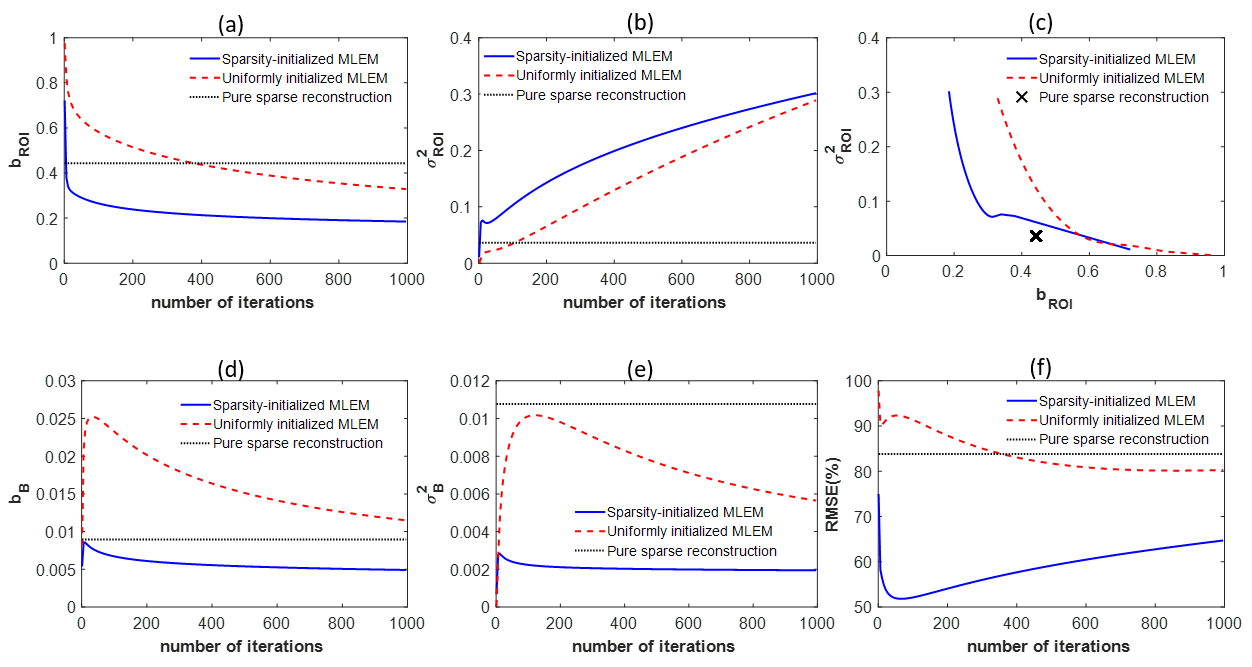}
\caption{Quantitative results of different reconstruction methods as functions of iteration number for cube phantom. (a) Plot of ROI bias vs. number of iterations. (b) Plot of ROI spatial variance vs. number of iterations. (c) Plot of ROI spatial variance vs. ROI bias. (d) Plot of background bias vs. number of iterations. (e) Plot of background variance vs. number of iterations. (f) Plot of RMSE vs. number of iterations.}
\label{fig:cube_quant}
\end{figure}
\begin{equation}
\theta_{\textrm{ROI}}=\frac{1}{N_R}\sum_{r=1}^{N_R}{x_r},
\end{equation}
where $r$ denotes the $r^{th}$ voxel in the ROI, and $N_R$ is the number of voxels in the ROI. Similarly, the background mean, denoted by $\theta_{\textrm{B}}$, is defined as
\begin{equation}
\theta_{\textrm{B}}=\frac{1}{N_B}\sum_{b=1}^{N_B}{x_b},
\end{equation}
where $b$ denote the $b^{th}$ voxel in the background region, and $N_B$ is the number of voxels in the background.
Then the ROI absolute bias, denoted by $b_{\textrm{ROI}}$, was computed as:
\begin{equation}
b_{\textrm{ROI}}=\frac{1}{R}\sum_{k=1}^R|\theta_{\textrm{ROI},k}-\theta^{true}_{\textrm{ROI},k}|,
\end{equation}
where $k$ denotes the $k^{th}$ noise realization, $\theta^{true}_{\textrm{ROI},k}$ denotes the true uptake in the $k^{th}$ voxel in the ROI, and $R$ is the total number of noise realizations.
The background absolute bias, denoted by $b_{\textrm{B}}$, was computed as:
\begin{equation}
b_{\textrm{B}}=\frac{1}{R}\sum_{k=1}^R|\theta_{\textrm{B},k}-\theta^{true}_{\textrm{B},k}|,
\end{equation}
where $\theta^{true}_{\textrm{B},k}$ denotes the true uptake in the $k^{th}$ voxel in the background.
We also computed the spatial variance within the pixels in the ROI (denoted by $\sigma^2_{ROI}$) and in the background (denoted by $\sigma^2_{B}$) as follows:
\begin{equation}
\sigma^2_{ROI}=\frac{1}{R(N_R-1)}\sum_{k=1}^R{\sum_{r=1}^{N_R}{(x_{r, k}-\theta_{\textrm{ROI},k})^2}}.
\end{equation}

\begin{equation}
\sigma^2_{B}=\frac{1}{R(N_B-1)}\sum_{k=1}^R{\sum_{b=1}^{N_B}{(x_{b, k}-\theta_{\textrm{B},k})^2}}.
\end{equation}
The RMSE over the entire 3D image was computed as below:
\begin{equation}
\textrm{RMSE}=\frac{1}{R}\sum_{k=1}^R\sqrt{\frac{\sum_{i=1}^N{(x_{i, k}-x^{true}_{i, k})^2}}{\sum_{i=1}^N{(x^{true}_{i, k})^2}}}\times 100\%,
\end{equation}
where the subscript $k$ denotes the $k^{th}$ noise realization and the subscript $i$ denotes the $i^{th}$ voxel.
%where $i$ denotes $i_{th}$ voxel of the entire image. 
\begin{figure}
\centering\includegraphics[width=\textwidth]{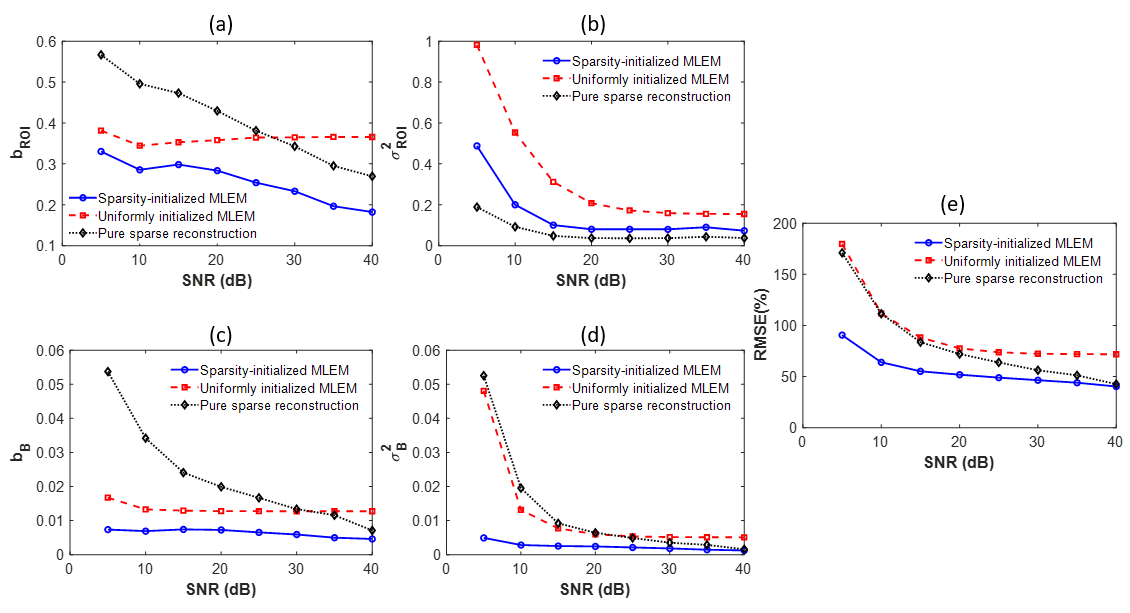}
\caption{Quantitative results of different reconstruction methods as functions of SNR for cube phantom. (a) Plot of ROI bias vs. SNR. (b) Plot of ROI variance vs. SNR. (c) Plot of background bias vs. SNR. (d) Plot of background variance vs. SNR. (e) Plot of RMSE vs. SNR.}
\label{fig:cube_SNR}
\end{figure}
In this and all the other experiments in this paper, $100$ noise realizations were used to compute the various figures of merit.
To study the sensitivity of our method to noise, experiments were conducted with SNR ranging from $5$ dB to $40$ dB, with step size of $5$ dB.
%\begin{figure}
%\centering\includegraphics[width=\textwidth]{MLEM_iteration_moby}
%\caption{Cross sections at $z=16\textrm{ mm}$ reconstructed by MLEM with different iteration number n for digital mouse phantom. SNR=$18$dB. The first row of reconstructed images are from MLEM with uniform initialization. The second row of reconstructed images are from MLEM with sparse initialization.}
%\label{fig:moby_crosec}
%\end{figure}

In the second set of experiments, we conducted simulation studies with a digital mouse phantom\cite{Segar2004}. Three fluorescence targets were placed in the mouse brain. Two of them had a radius of $0.8 \textrm{ mm}$ and the third had a radius of $1.2$ mm. The optical properties of the mouse head are listed in Table \ref{table1}. The whole brain was discretized into $2942$ voxels. $48$ sources and $51$ detectors were placed at the surface of the mouse head, as shown in Fig.~\ref{fig:moby_phan}(a). The cross section of the phantom at $z=16\textrm{mm}$ is shown in Fig.~\ref{fig:moby_phan}(b). 

First, $1000$ iterations were performed for MLEM with uniform and sparse initialization to study the effect of iteration number on MLEM performance. The SNR was set to $18$ dB. The truncation number $K$ was set to 120. Next, quantitative performance of pure sparse reconstruction, sparsity-initialized MLEM and uniformly initialized MLEM methods at different noise levels were evaluated. The SNR value ranged from $5$ dB to $40$ dB, with step size of $5$ dB.

The selection of truncation number plays an important role in the quality of the reconstructed image acquired from sparse reconstruction\cite{jin2014light,yao2015wide}. For this reason, we also studied the effect of truncation number on reconstruction results of pure sparse reconstruction method and the proposed sparsity-initialized MLEM method. To compare our proposed method and pure sparse reconstruction method, we conducted experiments with different truncation number $K$. $450$ iterations were used for MLEM with sparse initial estimate.
For quantitative study, RMSE was computed as a function of the truncation number. The experiments were conducted for two noise levels, namely SNR=40 dB and SNR=20 dB. 
\section{Results}
\subsection{Uniform cube phantom}
Fig.~\ref{fig:cube_crosec} shows cross sections reconstructed by MLEM with different iteration numbers. For sparsity-initialized MLEM, iteration number $n=0$ corresponds to the case of pure sparse reconstruction. The fluorescence intensity in all figures were normalized to the range of $[0, 1]$.
The computation time required by MLEM with different initializations for 1000 iterations is provided in Table \ref{table2}. It can be observed that sparsity-initialized MLEM is about 8 times faster than uniformly initialized MLEM.
\begin{table}[ht!]
\centering
\caption{Computation time required by MLEM for 1000 iterations}
\begin{tabular}{c | c}
\hline
Method & Time (s) \\
\hline
Sparsity-initialized MLEM & 5  \\
Uniformly initialized MLEM  & 39\\
\hline
\end{tabular}
\label{table2}
\end{table}
%We note that both initialization methods successfully resolve the two fluorescence target when iteration number is large. However, as iteration number increasing, it can be observed the spatial variance in region of interest (ROI) is also increased. To avoid this case, early stop is generally used in MLEM, which can be treated as a smooth regularizer. For MLEM with uniform initialization, a relative good reconstruction is achieved at iteration number between $500$ and $1000$, where the two fluorescence targets are separated and ROI variance is not too large. For MLEM with sparse initialization, the proper iteration number occurs between $10$ and $100$, where the background noise in the sparse reconstruction is suppressed and the ROI variance is not large.

Fig.~\ref{fig:cube_quant} shows the quantitative results as a function of iteration number. We observe from these plots that sparsity-initialized MLEM converges at a lower number of iterations. Sparsity-initialized MLEM has lower ROI bias, background bias, background spatial variance, and image RMSE. We also computed the variance of the mean ROI and the mean background uptakes, and found that these were much lower (less than $1\%$) compared to the bias. Thus, we do not show these results here. 
From Fig.~\ref{fig:cube_quant}(f), we notice that sparsity-initialized MLEM reached its lowest RMSE after only 50 iterations, but for uniformly initialized MLEM, the lowest RMSE was obtained after 800 iterations. Based on this result, we chose 50 iterations for sparsity-initialized MLEM and 800 iterations for uniformly initialized MLEM for the first set of experiments with different SNR values. The plots of quantitative results for the different reconstruction methods at different SNR values are shown in Fig.~\ref{fig:cube_SNR}. we again observe that sparsity-initialized MLEM leads to lower ROI bias, background bias, background spatial variance, and image RMSE for all noise levels.
\begin{figure}
\centering\includegraphics[width=\textwidth]{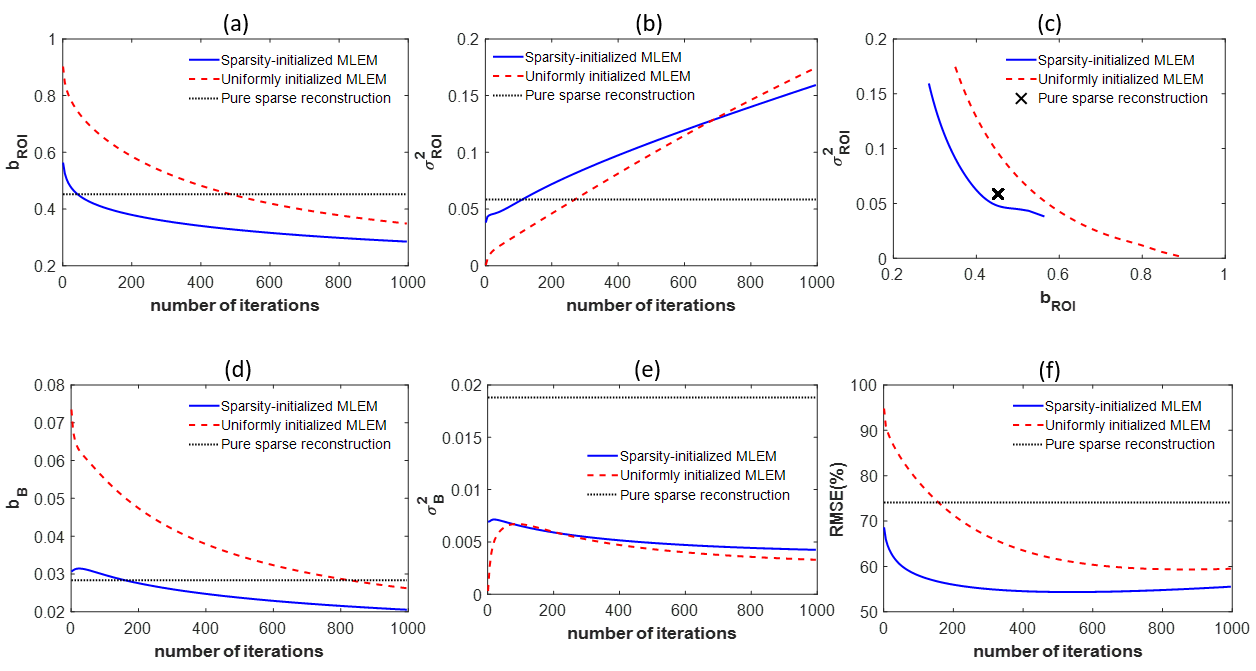}
\caption{Quantitative results of different reconstruction methods as functions of iteration number for digital mouse phantom. (a) Plot of ROI bias vs. number of iterations. (b) Plot of ROI spatial variance vs. number of iterations. (c) Plot of ROI spatial variance vs. ROI bias. (d) Plot of background bias vs. number of iterations. (e) Plot of background variance vs. number of iterations. (f) Plot of RMSE vs. number of iterations.}
\label{fig:moby_quant}
\end{figure}
\subsection{Digital mouse phantom}
%Fig. \ref{fig:moby_crosec} shows corss sections of reconstructed images of digital mouse phantom by uniform and sparsity-initialized MLEM with different iteration number $n$. The fluorescence intensity in all figures were normalized to the range of $[0, 1]$.
Fig.~\ref{fig:moby_quant} shows quantitative performance of different reconstruction methods as a function of iteration number. We observe that sparsity-initialized MLEM achieves lower ROI bias, background bias and RMSE. It was observed that for the sparsity-initialized MLEM and uniformly initialized MLEM, 450 and 900 iterations yielded the minimum RMSE. Thus, these values were chosen for the two methods for subsequent experiments.
Quantitative performance of different reconstruction methods at different noise levels is shown in Fig.~\ref{fig:moby_SNR}. The sparsity-initialized MLEM method shows better performance for ROI bias, background bias and RMSE compared to the other two methods. 

Fig.~\ref{fig:moby_crosecprecond} shows the cross sections reconstructed by pure sparse reconstruction and MLEM with sparse initial estimate for different truncation number. From Fig.~\ref{fig:moby_crosecprecond}, we notice that for small truncation number, pure sparse reconstruction generates blurry images. As truncation number increases, the resolution improves, but the background noise also increases due to the amplification of noise during preconditioning. For truncation number larger than 550, the signal is totally overwhelmed by the noise. As a comparison, the proposed method is able to largely reduce the background noise as truncation number increases. 
The RMSE as a function of truncation number is plotted in Fig.~\ref{fig:moby_prec}. The sparsity-initialized MLEM leads to lower RMSE for both noise levels.
\section{Discussion}
\begin{figure}
\centering\includegraphics[width=\textwidth]{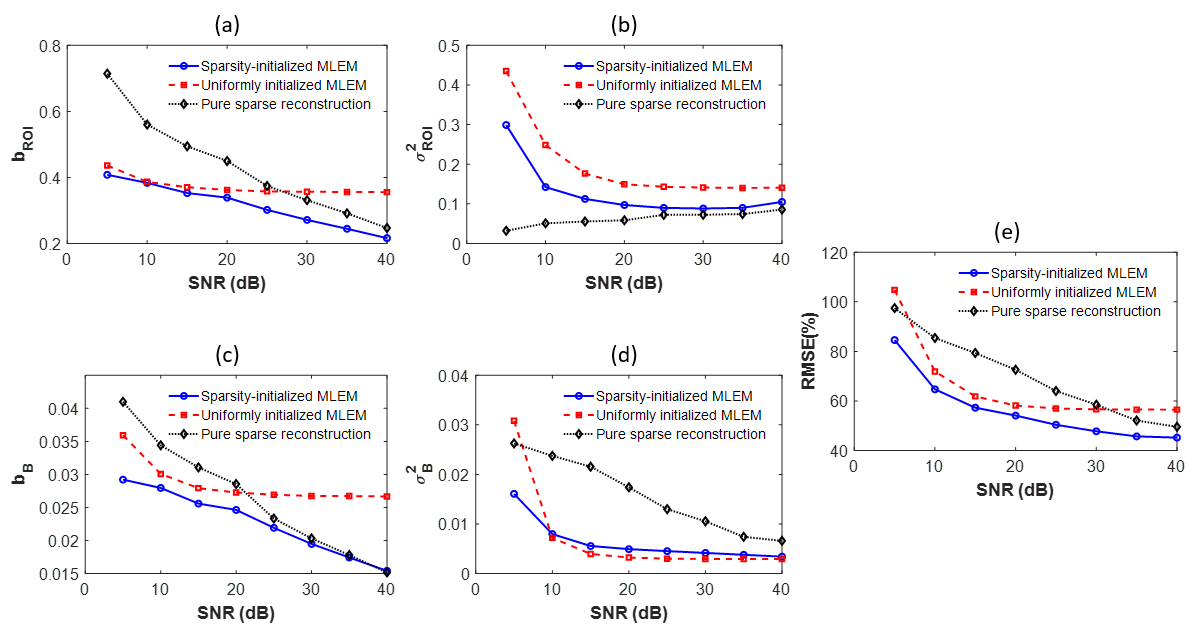}
\caption{Quantitative results of different reconstruction methods as functions of SNR for digital mouse phantom. (a) Plot of ROI bias vs. SNR. (b) Plot of ROI variance vs. SNR. (c) Plot of background bias vs. SNR. (d) Plot of background variance vs. SNR. (e) Plot of RMSE vs. SNR.}
\label{fig:moby_SNR}
\end{figure}

In this paper, we have proposed an MLEM-based technique to reconstruct the fluorescence distribution from FMT data. In our framework, the initial estimate for the MLEM algorithm is derived from a sparse reconstruction method. Often an uniform initial estimate is used with MLEM-based techniques, but here we observe that a sparsity-initialized technique yields several advantages compared to uniformly initialized MLEM. First, sparsity-initialized MLEM has faster convergence speed. From Table \ref{table2}, Fig.~\ref{fig:cube_crosec}, Fig.~\ref{fig:cube_quant}(a) and Fig.~\ref{fig:moby_quant}(a), we observe that sparse initial estimate speeds up the convergence by both shortening the computation time for each iteration and requiring fewer iterations for convergence. In addition, sparsity-initialized MLEM also provides improved quantitative performance in ROI bias, background bias, ROI spatial variance, RMSE, and bias-variance trade-off compared to uniformly initialized MLEM, as shown in Fig.~\ref{fig:cube_quant}-\ref{fig:moby_SNR}. Further, while results in both the cube phantom and the digital mouse phantom experiments indicate that the proposed method leads to higher ROI spatial variance compared to uniformly initialized MLEM for the same number of iterations, Fig.~\ref{fig:cube_quant}(c) and Fig.~\ref{fig:moby_quant}(c) show that the proposed method still provides better bias-variance trade-off compared to MLEM with uniform initial estimate. Further, sparsity-initialized MLEM often requires fewer iterations, which enables it to provide lower ROI spatial variance compared to uniformly initialized MLEM, as we observe from Fig.~\ref{fig:cube_SNR}(b) and Fig.~\ref{fig:moby_SNR}(b).

\begin{figure}
\centering\includegraphics[width=\textwidth]{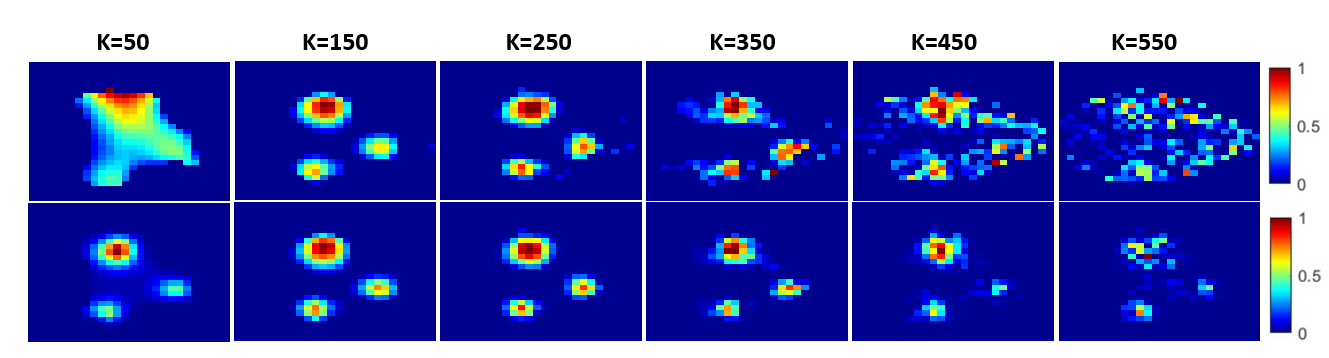}
\caption{Cross sections of fluorescence target reconstructed with pure sparse reconstruction method for the top row and the proposed method for the bottom row with different truncation number $K$ for digital mouse phantom for SNR=$40$dB.}
\label{fig:moby_crosecprecond}
\end{figure}

We also observe that sparsity-initialized MLEM provides advantages over pure sparse reconstruction method. From Fig.~\ref{fig:moby_crosecprecond}, we notice that sparsity-initialized MLEM is less sensitive to the choice of truncation number. For pure sparse reconstruction method, when the truncation number is small, the reconstructed image is blurry. As truncation number increases, image resolution is improved, but the noise in the background region is also increased due to the noise amplification during preconditioning. On the other hand, for small truncation number, sparsity-initialized MLEM is able to improve the resolution compared to pure sparse reconstruction method. For large truncation number, sparsity-initialized MLEM reduces noise in the background. These properties make MLEM with sparse initial estimate more robust to the choice of truncation number compared to pure sparse reconstruction method. The plots of RMSE vs. truncation number in Fig.~\ref{fig:moby_prec} also demonstrate this point. Sparsity-initialized MLEM also improves quantitative performance of reconstructed images compared to pure sparse reconstruction method. Fig. \ref{fig:cube_SNR} and Fig. \ref{fig:moby_SNR} indicate this for different SNR values. Apart from improved background bias and spatial variance due to the reduction of background noise, sparsity-initialized MLEM also reduces the ROI bias compared to pure sparse reconstruction method, especially at low SNR value. At low SNR value, small truncation number is preferred to avoid noise amplification, which results in only a small number of measurements used for reconstruction. Small truncation number not only generates blurry images, as we discussed previously, but also causes severe bias in the reconstruction results. For example, for SNR$=5\textrm{ dB}$, we observe that pure sparse reconstruction generated $71\%$ ROI bias in the cube phantom experiments and $57\%$ ROI bias in the digital mouse phantom experiments. As a comparison, MLEM uses the original system matrix and detector measurements for reconstruction, enabling it to compensate for the bias in the image, which reduced the ROI bias to $40\%$ in the cube phantom experiments and $33\%$ in the digital mouse phantom experiments. 

\begin{figure}
\centering\includegraphics[width=0.8\textwidth]{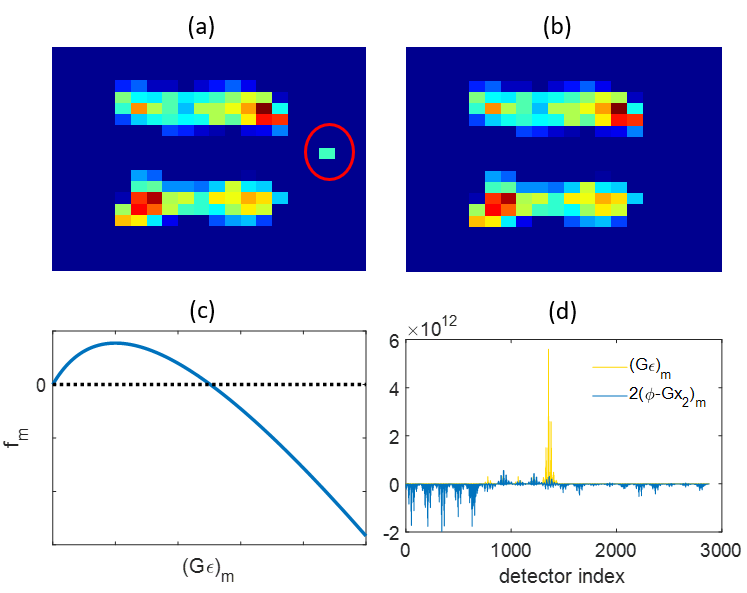}
\caption{(a) Image with background noise. The noise spot in the background is marked with red circle. (b) Image without background noise. (c) Plot of $f_m$. (d)Plot of $(\G\epsilon)_m$ and $2(\phi-\G\x_2)_m$}
\label{fig:discuss}
\end{figure}

We have observed, for example in Fig.~\ref{fig:cube_crosec}, that sparsity-initialized MLEM is able to suppress the noise in the background region that is present in the sparse initial estimate. To explain this observation, here we provide a theoretical justification. For a set of detector measurements denoted by $\PHI$, consider two reconstructed images $\x_1$ and $\x_2$, where $\x_1$ is image with noise in the background region (referred to as background noise), and $\x_2$ is image that does not contain this background noise, as shown in Fig.~\ref{fig:discuss}(a) and (b), respectively. We denote the background noise as $\EPSILON=\x_1-\x_2$, where $\epsilon_n\geq0$ for all $n$. 
Before we proceed further, we introduce the concept of KL distance. This distance measures how two probability distributions diverge from another. It is known that MLEM attempts to find an estimate that minimizes the Kullback-Leibler (KL) distance between the measured data $\PHI$ and the data predicted by an estimate $\G\x$. Thus, our objective is to assess whether the KL distance of $\x_2$ is less than $\x_1$, which would explain why MLEM would yield a solution $\x_2$ in comparison to $\x_1$. 
For $\x_1$, the KL distance is:
\begin{figure}
\centering
\subfloat[]{\includegraphics[width=0.45\textwidth]{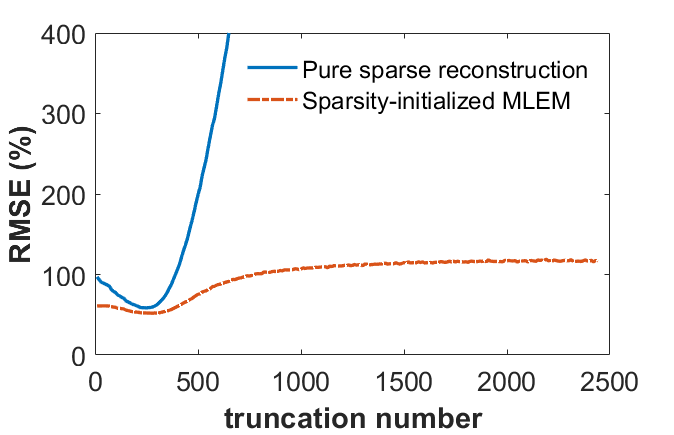}} \hspace{0.1 cm}
\subfloat[]{\includegraphics[width=0.45\textwidth]{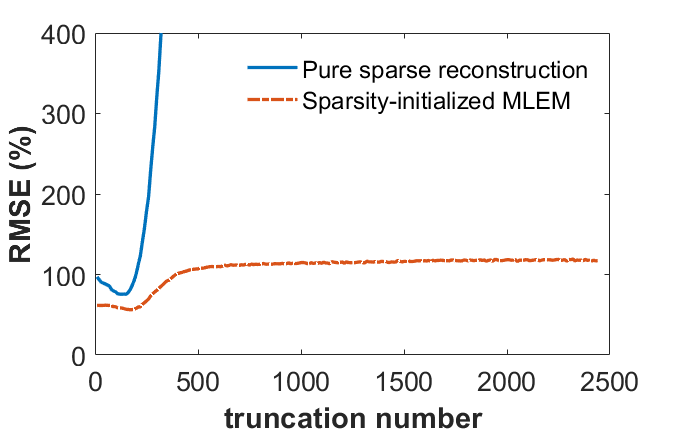}} \\
\caption{Plot of RMSE vs. truncation number for pure sparse reconstruction method and the proposed reconstruction method for different noise levels. (a) Plot of RMSE vs. truncation number for SNR=$40$ dB. (b) Plot of RMSE vs. truncation number for SNR=$20$ dB. }
\label{fig:moby_prec}
\end{figure} 
\begin{equation}
D_{KL,1}(\PHI,\G\x_1)=\sum_m\left\{(\G\x_1)_m-\phi_m+\phi_m\ln{\frac{\phi_m}{(\G\x_1)_m}}\right\}.
\end{equation}
For $\x_2$, the KL distance is:
\begin{equation}
D_{KL,2}(\PHI,\G\x_2)=\sum_m\left\{(\G\x_2)_m-\phi_m+\phi_m\ln{\frac{\phi_m}{(\G\x_2)_m}}\right\}.
\end{equation}
Then the difference is:
\begin{align}
\Delta D_{KL}&=D_{KL,2}(\PHI,\G\x_2)-D_{KL,1}(\PHI,\G\x_1)\nonumber\\
%&=\sum_m\{-(\G\EPSILON)_m+\BPHI_m\ln{\frac{(\G\x_1)_m}{(\G\x_2)_m}}\}\\
&=\sum_m\left\{-(\G\EPSILON)_m+\phi_m\ln{\left[1+\frac{(\G\EPSILON)_m}{(\G\x_2)_m}\right]}\right\}.
\end{align}
We denote $f_m((\G\EPSILON)_m)=-(\G\EPSILON)_m+\phi_m\ln{\left[1+\frac{(\G\EPSILON)_m}{(\G\x_2)_m}\right]}$. If $\phi_m\leq(\G\x_2)_m$, $f_m\leq 0$ since $f_m$ is monotonically decreasing for $(\G\EPSILON)_m\leq 0$ and $f_m(0)=0$. If $\phi_m>(\G\x_2)_m$, the plot of $f_m$ is shown in Fig.~\ref{fig:discuss}(c). 
To estimate the zeros of $f_m$, we use second order Taylor expansion to approximate $f_m$, which gives:
\begin{equation}
f_m\approx -(\G\EPSILON)_m+\phi_m\left[\frac{(\G\EPSILON)_m}{(\G\x_2)_m}-\frac{1}{2}\left(\frac{(\G\EPSILON)_m}{(\G\x_2)_m}\right)^2 \right].
\end{equation} 
Let $f_m=0$, we get 
\begin{equation}
(\G\EPSILON)_{m,1}=0,
\end{equation}
and 
\begin{align}
(\G\EPSILON)_{m,2}&=\frac{2(\G\x_2)_m[\phi_m-(\G\x_2)_m]}{\phi_m}\nonumber\\
&\leq2[\phi_m-(\G\x_2)_m],
\end{align}
where the inequality comes from the fact that $\phi_m>(\G\x_2)_m$. Also, note that the function $f_m$ has its maxima at $(\G\epsilon)_m=\phi_m-(\G\x_2)_m$. Thus, if the detector response to the noise spot has similar pattern as $\phi_m-(\G\x_2)_m$, $f_m$ will be close to its maximum for most detector index $m$. This provides a higher chance that $\Delta D_{KL}>0$, which means MLEM is more likely to update towards noisy image. This is the case for the noise close to ROI. On the other hand, for noise spot in background region, the detector response to noise spot will have a very different pattern compared to $\phi_m-(\G\x_2)_m$, as shown in Fig.~\ref{fig:discuss}(d). For detector index $m$ where $\phi_m-(\G\x_2)_m>0$, $(\G\epsilon)_m$ is either close to $0$ or too large. This results in makes $f_m$ either close to zero or have a negative value. In this case, it has higher chance that $\Delta D_{KL}<0$, meaning MLEM tends to update towards results without the noise spot. 

The noise model in FMT is often assumed to be Gaussian\cite{Zhao:14,ye2014fast,Han:10,dutta2012joint,shi2013greedy,yao2015wide}. In very few cases is the Poisson noise model applied\cite{7351233}. 
Gaussian noise model is a good approximation when SNR is high, i.e. sufficient number of photons are detected. However, in some applications, the SNR value might be low, such as in brain imaging\cite{Raymond:09}, dynamic FMT\cite{liu2011unmixing} and early-photon FMT\cite{leblond2009early}. Our results demonstrate that incorporating the Poisson noise model is especially valuable in these scenarios. More specifically, the pure sparse reconstruction method was formulated based on Gaussian noise model, while the proposed method incorporated both the sparsity information and Poisson noise model. We observe that the performance of the proposed method improves in comparison to the pure sparse reconstruction method as the SNR value decreases, and the proposed method is substantially more reliable at low SNR values. This shows the importance of accurately modeling Poisson noise for applications of FMT when insufficient number of photons are detected.

In this work, we only considered the case where the background uptake of fluorescence distribution is zero. While this is a common assumption in FMT studies\cite{Zhao:14,ye2014fast,Han:10,dutta2012joint,shi2013greedy,jin2014light,yao2015wide}, it is possible that the background uptake is non-zero. Exploring the performance of the proposed method for this task would be an important future direction. The proposed method has been validated with extensive simulation experiments. Evaluating the performance of the method with physical phantom and in \textit{vivo} animal experiments is another important direction of research. Finally, we used the MC-based method to model photon propagation to obtain the Greens function in this work. However, there have been several analytical methods proposed for modeling light transport\cite{Lehtikangas:12,Tarvainen:05,mohan2011variable,Jha:12,jha2012three,Jha:17}. These methods can also be used to obtain an expression for the Green's function. Analytical methods offer the advantage that they might be less sensitive to photon noise. Thus, implementing this reconstruction method using the analytical approaches is another important research direction. 
\section{Conclusion}
We have presented a reconstruction framework for FMT involving sparsity-initialized MLEM. Simulation experiments on cubic digital mouse phantoms demonstrate that the proposed method yields improved qualitative and quantitative performance compared to uniformly initialized MLEM as well as sparsity-initialized MLEM techniques. Further, compared to uniformly initialized MLEM, the proposed method is faster to execute, overcoming another barrier to application of MLEM technique for optical tomography. Moreover, compared to pure sparse reconstruction, the proposed method is more robust to noise amplification. We have also provided theoretical justification for the ability of the proposed method to reduce noise in the background region. Overall, this paper provides strong evidence that the proposed sparsity initialized MLEM-based reconstruction framework is feasible and advantageous for reconstruction in FMT imaging systems.
\section{Funding}
NIH BRAIN Initiative Award R24 MH106083.
\section{Acknowledgments}
The authors thank Drs. Eric Frey and Jin Kang for helpful discussions.
\section{Disclosures}
Dean Wong acknowledges contract work with Lilly, Lundbeck, Intracellular, Five Eleven Pharma, Roche and Dart pharmaceuticals.
\end{document}